\documentclass[12pt]{article}%
\usepackage[sort]{cite}
\usepackage{graphicx}
\usepackage{multicol}
\usepackage{amsfonts}
\usepackage{amssymb}
\usepackage{amsmath}
\usepackage{heck}
\usepackage{afterpage}
\usepackage{setspace}
\usepackage{verbatim}
\usepackage{color}
\usepackage{longtable}
\usepackage{standalone}
\usepackage{float}
\usepackage{subcaption}
\usepackage{epsfig}
\usepackage{adjustbox}
\usepackage{tikz}
\usepackage{soul}
\usepackage[margin=1in]{geometry}
\usepackage{titletoc}
\usepackage{hyperref}%
\providecommand{\U}[1]{\protect\rule{.1in}{.1in}}
\pdfoutput=1
\newsavebox{\mysavebox}

\hypersetup{colorlinks,citecolor=black,filecolor=black,linkcolor=black,urlcolor=black,pdftex}
\numberwithin{equation}{section}

\newcommand{\ba}{\begin{eqnarray}}
\newcommand{\ea}{\end{eqnarray}}

\newcommand{\be}{\begin{equation}}
\newcommand{\ee}{\end{equation}}

\usepackage{tikz}
\usetikzlibrary{arrows}
\usetikzlibrary{arrows.meta}
\usetikzlibrary{shapes.geometric,calc,arrows, positioning,shapes.misc,decorations.markings}
\tikzset{
  big arrow/.style={
    decoration={markings,mark=at position 1 with {\arrow[scale=2,#1]{>}}},
    postaction={decorate},
    shorten >=0.4pt},
  big arrow/.default=black}

\pgfdeclarelayer{edgelayer}
\pgfdeclarelayer{nodelayer}
\pgfsetlayers{edgelayer,nodelayer,main}
\tikzstyle{none}=[inner sep=0pt]

\tikzstyle{NodeCross}=[draw, shape=circle, cross out, inner sep=0pt, minimum size=6pt,line width=0.25mm]
\tikzstyle{Circle}=[draw, shape=circle, black, fill=black, inner sep=0pt, minimum size=6pt]
\tikzstyle{Star}=[draw, shape=star, fill=black, star points=8, inner sep=0pt, minimum size=8pt]
\tikzstyle{CircleRed}=[draw, shape=circle, black, fill=red, inner sep=0pt, minimum size=4pt]
\tikzstyle{StarP}=[draw={rgb,255: red,128; green,0; blue,128}, shape=star, fill={rgb,256: red,128; green,0; blue,128}, star points=8, inner sep=0pt, minimum size=12pt]

\tikzstyle{DashedLine}=[-, densely dashed, line width=0.25mm]
\tikzstyle{DottedLine}=[-, dotted, line width=0.25mm]
\tikzstyle{ThickLine}=[-, line width=0.25mm]
\tikzstyle{ArrowLineRight}=[-, -{Stealth[scale=1.75]}, line width=0.1mm, scale=5]
\tikzstyle{ArrowLineRed}=[-, draw={rgb,255: red,191; green,0; blue,0}, -{Stealth[scale=1.75]}, line width=0.1mm, scale=5]
\tikzstyle{RedLine}=[-, draw={rgb,255: red,191; green,0; blue,0}, fill=none, line width=0.25mm]
\tikzstyle{DashedLineThin}=[-, densely dashed, line width=0.125mm, fill=none, draw=black]
\tikzstyle{DottedRed}=[-, dotted, draw={rgb,255: red,191; green,0; blue,0}, fill=none, line width=0.25mm]
\tikzstyle{DashedRed}=[-, densely dashed, draw={rgb,255: red,191; green,0; blue,0}, fill=none, line width=0.25mm]
\tikzstyle{BlueLine}=[-, draw={rgb,255: red,0; green,0; blue,191}, fill=none, line width=0.25mm]

\begin{document}

\date{September 2022}

\title{The Branes Behind \\[4mm] Generalized Symmetry Operators}

\institution{PENN}{\centerline{$^{1}$Department of Physics and Astronomy, University of Pennsylvania, Philadelphia, PA 19104, USA}}
\institution{PENNmath}{\centerline{$^{2}$Department of Mathematics, University of Pennsylvania, Philadelphia, PA 19104, USA}}

\authors{
Jonathan J. Heckman\worksat{\PENN,\PENNmath}\footnote{e-mail: \texttt{jheckman@sas.upenn.edu}},
Max H\"ubner\worksat{\PENN}\footnote{e-mail: \texttt{hmax@sas.upenn.edu}},\\[4mm]
Ethan Torres\worksat{\PENN}\footnote{e-mail: \texttt{emtorres@sas.upenn.edu}}, and
Hao Y. Zhang\worksat{\PENN}\footnote{e-mail: \texttt{zhangphy@sas.upenn.edu}}
}

\abstract{The modern approach to $m$-form global
symmetries in a $d$-dimensional quantum field theory (QFT)
entails specifying dimension $d-m-1$ topological generalized symmetry operators which
non-trivially link with $m$-dimensional defect operators. In QFTs engineered
via string constructions on a non-compact geometry $X$,
these defects descend from branes wrapped on non-compact cycles which extend from a localized
source / singularity to the boundary $\partial X$.
The generalized symmetry operators which link with these defects arise from magnetic dual branes
wrapped on cycles in $\partial X$. This provides a systematic
way to read off various properties of such topological operators,
including their worldvolume topological field theories, and the
resulting fusion rules. We illustrate these general features in
the context of 6D superconformal field theories, where we use the F-theory
realization of these theories to read off the worldvolume theory on the generalized
symmetry operators. Defects of dimension 3 which are charged under a suitable 3-form symmetry
detect a non-invertible fusion rule for these operators.
We also sketch how similar considerations hold for related systems.}

\maketitle

\enlargethispage{\baselineskip}

\setcounter{tocdepth}{2}

\section{Introduction}

One of the important recent advances in the study of quantum field theory
(QFT) has been the appreciation that generalized symmetries can often be
better understood in terms of corresponding topological operators
\cite{Gaiotto:2014kfa}. For a $d$-dimensional QFT, an $m$-form symmetry
naturally acts on $m$-dimensional defects. There is a corresponding
dimension $d-m-1$ generalized symmetry operator which is topological, i.e., it is unchanged by local perturbations to
its shape \cite{Gaiotto:2014kfa, Gaiotto:2010be,Kapustin:2013qsa,Kapustin:2013uxa,Aharony:2013hda}.
This includes higher-form symmetries, their entwinement via higher-groups,
as well as more general categorical structures.\footnote{For recent work in
this direction, see e.g.,
\cite{Gaiotto:2014kfa,Gaiotto:2010be,Kapustin:2013qsa,Kapustin:2013uxa,Aharony:2013hda,
DelZotto:2015isa,Sharpe:2015mja, Heckman:2017uxe, Tachikawa:2017gyf,
Cordova:2018cvg,Benini:2018reh,Hsin:2018vcg,Wan:2018bns,
Thorngren:2019iar,GarciaEtxebarria:2019caf,Eckhard:2019jgg,Wan:2019soo,Bergman:2020ifi,Morrison:2020ool,
Albertini:2020mdx,Hsin:2020nts,Bah:2020uev,DelZotto:2020esg,Hason:2020yqf,Bhardwaj:2020phs,
Apruzzi:2020zot,Cordova:2020tij,Thorngren:2020aph,DelZotto:2020sop,BenettiGenolini:2020doj,
Yu:2020twi,Bhardwaj:2020ymp,DeWolfe:2020uzb,Gukov:2020btk,Iqbal:2020lrt,Hidaka:2020izy,
Brennan:2020ehu,Komargodski:2020mxz,Closset:2020afy,Thorngren:2020yht,Closset:2020scj,
Bhardwaj:2021pfz,Nguyen:2021naa,Heidenreich:2021xpr,Apruzzi:2021phx,Apruzzi:2021vcu,
Hosseini:2021ged,Cvetic:2021sxm,Buican:2021xhs,Bhardwaj:2021zrt,Iqbal:2021rkn,Braun:2021sex,
Cvetic:2021maf,Closset:2021lhd,Thorngren:2021yso,Sharpe:2021srf,Bhardwaj:2021wif,Hidaka:2021mml,
Lee:2021obi,Lee:2021crt,Hidaka:2021kkf,Koide:2021zxj,Apruzzi:2021mlh,Kaidi:2021xfk,Choi:2021kmx,
Bah:2021brs,Gukov:2021swm,Closset:2021lwy,Yu:2021zmu,Apruzzi:2021nmk,Beratto:2021xmn,Bhardwaj:2021mzl,
Debray:2021vob, Wang:2021vki,
Cvetic:2022uuu,DelZotto:2022fnw,Cvetic:2022imb,DelZotto:2022joo,
DelZotto:2022ras,Bhardwaj:2022yxj,Hayashi:2022fkw,
Kaidi:2022uux,Roumpedakis:2022aik,Choi:2022jqy,
Choi:2022zal,Arias-Tamargo:2022nlf,Cordova:2022ieu, Bhardwaj:2022dyt,
Benedetti:2022zbb, Bhardwaj:2022scy,Antinucci:2022eat,Carta:2022spy,
Apruzzi:2022dlm, Heckman:2022suy, Baume:2022cot, Choi:2022rfe,
Bhardwaj:2022lsg, Lin:2022xod, Bartsch:2022mpm, Apruzzi:2022rei,
GarciaEtxebarria:2022vzq, Cherman:2022eml}. For a recent overview of generalized symmetries,
see reference \cite{Cordova:2022ruw}.} One of the general aims in this
direction is to use such topological structures to gain access to
non-perturbative information on various QFTs. This is especially important in
the context of strongly coupled systems where one typically does not have
access to a practically useful Lagrangian description of the system.

In this vein, one of the lessons from recent work in stringy realizations of QFT is
that there are large families of QFTs which do not have a (known) Lagrangian
description. This includes, for example, all 6D superconformal field theories
(SCFTs), as well as many compactifications of these theories.\footnote{See
\cite{Witten:1995zh,Strominger:1995ac,Seiberg:1996qx} for early examples, and
for recent work on the construction and study of such theories, see
\cite{Heckman:2015bfa,Tachikawa:2015wka,Heckman:2013pva,DelZotto:2014hpa, Heckman:2014qba, Intriligator:2014eaa,
Ohmori:2014pca, Ohmori:2014kda, DelZotto:2014fia,
Bhardwaj:2015xxa, DelZotto:2015isa, Bhardwaj:2015oru,
Cordova:2018cvg, Bhardwaj:2018jgp, Heckman:2018pqx, Bhardwaj:2019hhd,
Bergman:2020bvi,Baume:2020ure, Heckman:2020otd,Baume:2021qho, Distler:2022yse,
Heckman:2022suy} as well as \cite{Heckman:2018jxk, Argyres:2022mnu} for recent
reviews.} More broadly, one can consider the QFT limit of \textit{any} string
background $X$, as obtained by decoupling gravity. In this context, it is
natural to expect that the extra-dimensional geometry directly encodes these
generalized symmetries.

This expectation is, to a large extent, borne out by the explicit construction
of the defect operators of these systems. In the stringy setting, we can
generate defects, i.e., non-dynamical objects with formally infinite tension
by wrapping branes on non-compact cycles of $X$. The resulting higher
symmetries act on these objects, but can be partially screened by dynamical
degrees of freedom wrapped on compact cycles of the geometry. This generalized
screening argument \`{a} la 't Hooft was used in \cite{DelZotto:2015isa}
to define the ``defect group'' of a 6D SCFT. As noted in
\cite{GarciaEtxebarria:2019caf, Albertini:2020mdx, Morrison:2020ool},
specifying a polarization of the defect group amounts to determining the
electric / magnetic higher-form symmetries of the system. This perspective has
by now been generalized in a number of directions, and has reached the stage
where there are explicit algorithms for reading off generalized symmetries for
a large number of geometries \cite{DelZotto:2015isa, Albertini:2020mdx, DelZotto:2020esg, Gukov:2020btk, DelZotto:2020sop, Apruzzi:2020zot, Bhardwaj:2020phs, Bhardwaj:2021pfz, Apruzzi:2021vcu, Agrawal:2015dbf, Cvetic:2021maf, Apruzzi:2021mlh,
Apruzzi:2021nmk, Tian:2021cif, Bhardwaj:2021mzl, DelZotto:2022fnw, Hubner:2022kxr, Cvetic:2022imb, DelZotto:2022joo, Heckman:2022suy}.

One of the puzzling features of these analyses is that the topological
operators of reference \cite{Gaiotto:2014kfa} are in some sense only
implicitly referenced in such stringy constructions. The absence of an explicit brane
realization of these symmetry topological operators makes it challenging to
access some features of generalized symmetries in these systems. For example, it is
well-known in various weakly coupled examples that generalized symmetry operators
can support a topological field theory, and that in the context of theories
with non-invertible symmetries, these can also produce a non-trivial fusion algebra.

In this note we present a general prescription for how to construct
topological operators in the context of geometric engineering. We mainly focus
on the tractable case of 2-form symmetries for 6D SCFTs and their
compactification, as engineered via F-theory backgrounds. In these cases, the
generalized symmetry operators arise from D3-branes wrapped on boundary torsional cycles.
We find that when the $SL(2,\mathbb{Z})$ bundle of the F-theory model is non-trivial, these models
generically have a non-invertible symmetry simply because the fusion algebra for the
generalized symmetry operators contains multiple summands. This is quite analogous
to what has been observed in the context of various field theoretic constructions
\cite{Thorngren:2019iar,Komargodski:2020mxz, Gaiotto:2020iye, Nguyen:2021naa, Heidenreich:2021xpr, Thorngren:2021yso, Agrawal:2015dbf, Robbins:2021ibx, Robbins:2021xce, Sharpe:2021srf, Koide:2021zxj, Huang:2021zvu,
Inamura:2021szw, Cherman:2021nox, Kaidi:2021xfk, Choi:2021kmx, Wang:2021vki, Bhardwaj:2022yxj,
Hayashi:2022fkw, Sharpe:2022ene, Choi:2022zal, Kaidi:2022uux, Choi:2022jqy, Cordova:2022ieu,
Bashmakov:2022jtl, Inamura:2022lun, Damia:2022bcd, Choi:2022rfe, Lin:2022dhv, Bartsch:2022mpm,
Lin:2022xod, Cherman:2022eml, Burbano:2021loy, Damia:2022rxw} as well as some recent holographic models \cite{Apruzzi:2022rei, GarciaEtxebarria:2022vzq}.

We emphasize, however, that the construction we present can be applied to
essentially any QFT which can be engineered via a string / M- / F-theory
compactification. We expect that experts may already be aware of various aspects of this construction, but as far as we are aware, the closest analog of our construction only appeared a few weeks ago in the context of some specific holographic constructions \cite{Apruzzi:2022rei,GarciaEtxebarria:2022vzq}.

\section{Branes and Generalized Symmetry Operators}\label{sec:branesandgensym}

Our interest will be in understanding the brane realization of generalized
symmetry operators. To frame the discussion to follow, let us first briefly
recall how defects are engineered in such systems. We begin by considering a QFT
engineered via a string / M-theory background of the form $\mathbb{R}%
^{d-1,1}\times X$ where $X$ is taken to be a non-compact
$D$-dimensional geometry ($d+D=10$ for a string background and $d+D=11$ for an
M-theory background). We get a QFT by introducing branes and / or localized
singularities at a common point of $X$. These singularities need not be
isolated, and can in principle extend all the way out to the boundary
$\partial X$. Gravity is decoupled because $X$ is non-compact.
This provides a general template for engineering a wide range of (typically
supersymmetric) QFTs.

We obtain supersymmetric defects by wrapping BPS branes on non-compact cycles
of $X$ which extend from the localized singularity out to the boundary. As
explained in \cite{DelZotto:2015isa, GarciaEtxebarria:2019caf, Morrison:2020ool,
Albertini:2020mdx} a screening argument \`{a} la 't Hooft
then tells us that there is a corresponding set of unscreened defects:
\begin{equation}
\mathbb{D=}\underset{m}{\mathbb{%
{\displaystyle\bigoplus}
}}\mathbb{D}^{(m)}\text{ \ \ with \ \ }\mathbb{D}^{(m)}=\underset{p-k=m-1}{%
{\displaystyle\bigoplus}
}\frac{H_{k}(X,\partial X)}{H_{k}(X)},
\end{equation}
where in the above, the superscript $m$ references an $m$-dimensional defect
acted on by an $m$-form symmetry, we sum over supersymmetric $p$-branes and $k$ is a cycle
dimension. Specifying a polarization of $\mathbb{D}$ picks an electric /
magnetic basis of operators and also dictates the higher-form symmetries via
the Pontryagin dual. In the $d$-dimensional QFT, a $p$-brane wrapped on a
$k$-cycle will fill out $p+1-k$ spacetime dimensions. We indicate this by
saying that the brane fills the space:\footnote{We reserve untilded quantities for the
generalized symmetry operators.}
\begin{equation}
\widetilde{S}_{p+1}=\widetilde{M}_{p+1-k}\times\widetilde{\Sigma}_{k}= \widetilde{M}_{m}\times \widetilde{\Sigma}_{k},
\end{equation}
where $\widetilde{\Sigma}_{k} = \mathrm{Cone}(\widetilde{\gamma}_{k-1})$ is the cone generated by extending the
boundary cycle $\widetilde{\gamma}_{k-1} \in H_{k-1}(\partial X)$ from infinity to the tip of the cone.
A final comment is that the torsional factors of $\mathbb{D}^{(m)}$ will define discrete
higher-form symmetries. Non-torsional generators instead label continuous symmetries.

In the $d$-dimensional QFT, the appearance of an $m=(p+1-k)$-dimensional defect
implicitly means there are also corresponding operators $\mathcal{O}_{m}$
with support on an $m$-dimensional subspace. The $m$-form symmetry acts on the defects and operators by
passing these operators through a topological operator $\mathcal{U}(M_n)$ with support on an
$n$-dimensional subspace. To link with the defect, we have:%
\begin{equation}
m+n=d-1.
\end{equation}

\begin{figure}[t!]
\begin{center}
\includegraphics[scale = 1.0, trim = {0cm 0.0cm 0cm 0.0cm}]{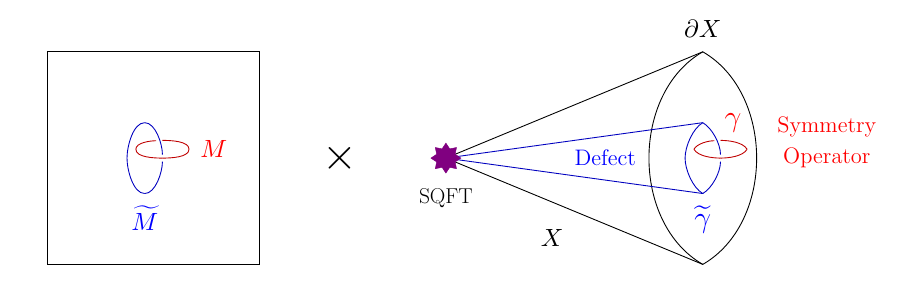}
\caption{Depiction of a generalized symmetry operator realized via a wrapped brane. We consider a
supersymmetric QFT (SQFT) localized at the tip of a conical geometry $X$. At the boundary $\partial X$, we have
bounding cycles $\widetilde{\gamma}$ and $\gamma$ which link in $\partial X$. Wrapping a $p$-brane on the cone over $\widetilde{\gamma}$ yields a
defect for an $m$-form symmetry, and wrapping the magnetic dual $q$-brane on a linking cycle $\gamma$
realizes a generalized symmetry operator. The spacetime loci $\widetilde{M}$ of the $p$-brane and $M$ of the $q$-brane also link with each other. The wrapped $q$-brane is automatically topological since it sits at infinity in the radial geometry.}
\label{fig:TopOpLinking}
\end{center}
\end{figure}

In seeking out an extra-dimensional origin for these operators, we
first observe that the defect embeds in spacetime,
and extends along the radial direction which starts at the tip of the
singularity and goes all the way to the boundary $\partial X$, wrapping a boundary cycle of
$H_{k-1}(\partial X)$, namely it is specified by an element of $H_{k}(X, \partial X)$.
Our general proposal is that the topological operator
which links with this object is given by a magnetic dual brane which links with the original
brane in both $X$ as well as the spacetime $\mathbb{R}^{d-1,1}$. In
particular, we can wrap a $q$-brane on a cycle of the form:%
\begin{equation}
S_{q+1}= M_{q+1-(D-1-k)}\times \gamma_{D-1-k}%
= M_{n}\times \gamma_{D-1-k},
\end{equation}
where $\gamma_{D-1-k}$ is a cycle in $H_{D-1-k}(\partial X)$ and $M_{q+1-(D-1-k)}$ is a subspace
of the $d$-dimensional spacetime. Observe that in
$X$, the cycle does not fill the \textquotedblleft radial
direction\textquotedblright. Rather, it always \textquotedblleft sits at
infinity\textquotedblright. Now, for this to be a topological operator which
properly links with the defect, we also require:%
\begin{equation}
(p+1-k)+(q+1-(D-1-k))=d-1,
\end{equation}
or equivalently:%
\begin{equation}
p+q=D+d-4.
\end{equation}
But observe that this is just the requirement that in the full string / M-theory background,
our sought after $q$-brane is simply the magnetic dual $p$-brane!
See figure \ref{fig:TopOpLinking} for a depiction.

We now argue that wrapping a $q$-brane on the cycle at infinity
$S_{q+1} = M_{q+1 - (D-1-k)} \times \gamma_{D-1-k}$
can be viewed as inserting a topological operator for the $m$-form symmetry.
Along these lines, recall that for a (supersymmetric) $p$-brane, there is a corresponding
$(p+1)$-form potential $\widetilde{C}_{p+1}$ which couples to this object, and thus a
$(p+2)$-form field strength $\widetilde{F}_{p+2}$. In the full
higher-dimensional geometry, we also can speak of the dual field strength
$\ast \widetilde{F}_{p+2}= F_{q+2}$, as sourced by a magnetic dual $(q+1)$-form potential
$C_{q+1}$. The presence of a $p$-brane signals that there is a modified Bianchi identity:
\begin{equation}
d F_{q+2} = \delta_{\widetilde{S}_{p+1}},
\end{equation}
namely a delta function supported contribution to the flux. Of course, this is just indicating that we have
a suitable symmetry current which couples to the magnetic dual potential $C_{q+1}$. Following the general
reasoning in \cite{Gaiotto:2014kfa}, we conclude that we can represent our topological symmetry operator
by exponentiating the integral of $F_{q+2}$ over the $(q+2)$-chain $Z_{q+2} = M_{q+1 - (D-1-k)} \times \mathrm{Cone}(\gamma_{D-1-k})$:
\begin{equation}
\mathcal{U}(Z) = \exp \left( 2 \pi i \int_{Z} F_{q + 2} \right).
\end{equation}
In particular, if we wish to know the action of $\mathcal{U}(Z)$ on any correlation function,
we can simply insert it into the path integral of the $(D+d)$-dimensional supergravity theory.

On the other hand, since we are integrating $F_{q + 2}$ over a chain, we can reduce
this to the integral of $C_{q+1}$ over the
bounding geometry $S_{q+1} = M_{q+1 - (D-1-k)} \times \gamma_{D-1-k}$. Since this is an insertion in the path integral, we can view this is as just specifying a ``brane at infinity'', namely a $(q+1)$-brane wrapped on the prescribed cycle. Indeed, integrating $C_{q+1}$ over this cycle is just one of the worldvolume couplings of the $q$-brane action. Note also that by wrapping the $q$-brane on a linking cycle at infinity, it is automatically topological from the perspective of the $d$-dimensional QFT. Indeed, any localized fluctuation of such a brane is completely decoupled from the dynamics of the field theory localized at the singularity. The only remaining structure which could in principle persist
is purely topological in nature.\footnote{Another way to arrive at the same conclusion is
to consider localized fluctuations from the singularity. Any correlation function involving
operators of the theory will be---up to topological couplings---completely decoupled from the ``brane at infinity''.
Thus, the only possible remnant of the brane at infinity on the localized dynamics could be topological in nature.}

Summarizing, our proposal is that for a wrapped $p$-brane which produces a defect, the corresponding
generalized symmetry operators which act on these defects are realized
by magnetic dual $q$-branes wrapped on linking cycles of the geometry.
This is compatible with the holographic discussion considered a few weeks ago in \cite{Apruzzi:2022rei, GarciaEtxebarria:2022vzq},
which considers the case of QFTs engineered via D3-brane probes of appropriate singularities. In that setting, the near horizon geometry is of the form $AdS_5 \times Y$ where $Y$ can be viewed as the asymptotic geometry $\partial X = Y$ probed by the D3-brane. Indeed, as noted in \cite{GarciaEtxebarria:2022vzq}, defects arise from branes which fill the radial direction of the $AdS_5$, while the symmetry operators arise from branes wrapped on a cycle of $Y$ and sitting at a point of the conformal boundary $\partial AdS_5$. It is important to emphasize that precisely because we are dealing with a conformal boundary the construction presented in the holographic setting is indeed compatible with the perspective developed here.

Observe that we can also read off the corresponding topological field theory
(TFT) localized on this symmetry operator. Starting from
$S_{q+1} = M_{q+1-(D-1-k)}\times \gamma_{D-1-k}$, we consider the topological
couplings on the worldvolume theory of our $q$-brane. Roughly speaking, we can
integrate this theory along $\gamma_{D-1-k}$ and arrive at a TFT\ on
$M_{n} = M_{q+1-(D-1-k)}$. To see this procedure through from start to finish, then,
we need to know the topological couplings on the original brane, as well as a
technique to dimensionally reduce along $\gamma_{D-1-k}$.

As a further abstraction, now that we have a method for realizing generalized symmetry operators, we can in principle
just consider branes wrapped on torsional cycles ``at infinity''. In particular, there is a
priori no need for there to exist explicit defects of the appropriate codimension which link
with these branes.\footnote{This, for example, happens in various 3D Chern-Simons-like theories with charge conjugation,
i.e., there is a (non-invertible) 0-form symmetry which acts on no local operators,
but line operators do transform non-trivially
in passing through the wall (see e.g., \cite{Seiberg:2020bhn, Choi:2022zal}).}

\section{Example: 2-Form Symmetries of 6D SCFTs}

To illustrate the above considerations, we now show how this works in practice for
(discrete) 2-form symmetries of 6D\ SCFTs. All known 6D SCFTs can be engineered
via F-theory on an elliptically fibered Calabi-Yau threefold with base $\mathcal{B}$ such that the threefold
has a canonical singularity \cite{Heckman:2013pva, DelZotto:2014hpa, Heckman:2015bfa}.
In the SCFT\ limit, all of the bases take the form $\mathcal{B}=\mathbb{C}^{2}%
/\Gamma$ for $\Gamma$ an appropriate finite subgroup of $U(2)$ (see
\cite{Heckman:2013pva} for the classification of all such $\Gamma$).
The defect group for the 2-form symmetry is $\mathrm{Ab}[\Gamma]$,
the abelianization of $\Gamma$ (see reference \cite{DelZotto:2015isa}).
Some basic features of these orbifold singularities are summarized in table \ref{tab:defectgrps}.

\begin{table}
\begin{center}
\renewcommand{\arraystretch}{1.25}
\begin{tabular}{|| c | c |  c  ||}
\hline
   $\Gamma$ & $\mathbb{D}^{(2)}$ & $L_\Gamma$  \\ [0.5ex]
 \hline\hline $\mathbb{Z}_{N}$ & $\mathbb{Z}_{N}$ & $1/N$ \\
\hline $D_{2N}$ & $\mathbb{Z}_2\times \mathbb{Z}_2$ & $\frac{1}{2}\begin{pmatrix}
  N & N-1 \\
  N-1 & N
\end{pmatrix} $\\
\hline $ D_{2N+1}$ & $\mathbb{Z}_4$ & $\frac{2N-1}{4}$ \\
\hline $2T$ & $\mathbb{Z}_3$ & $1/3$\\
\hline $2O$ & $\mathbb{Z}_2$ & $1/2$\\
\hline $2I$ & 1 & 1\\
\hline $\mathbb{Z}_p(q)$ & $\mathbb{Z}_p$ & $-q/p$\\
\hline $D_{p+q,q}$ & $\begin{array}{c}
   \mathbb{Z}_{2p}\times \mathbb{Z}_2~ \textnormal{($q$ even)} \\
      \mathbb{Z}_{2p} ~ \textnormal{($q$ odd)}
\end{array} $
& (\textnormal{See main text})\\
\hline
\end{tabular}
\end{center}
 \caption{In the left column we list out all of the families of finite subgroups of $U(2)$ associated to 6D SCFTs. Here $D_{k}$ means the dicyclic groups of order $2k$, and $2T$, $2O$, and $2I$ denote the binary tetrahedral, octahedral, and icosahedral groups respectively. $\mathbb{Z}_p(q)$ denotes a $\mathbb{Z}_p$ subgroup of $U(2)$ generated by an action $(z_1,z_2)\mapsto (\zeta_p z_1,\zeta^q_p z_2)$ ($p$ and $q$ coprime). Finally, $D_{p+q,q}$ is a $U(2)$ subgroup which generalizes the dicyclic group (see \cite{Heckman:2013pva, DelZotto:2015isa} and references therein for more details). }\label{tab:defectgrps}
\end{table}

It is helpful to decompose the base geometry as a fibration $S^{3}%
/\Gamma\rightarrow\mathcal{B}\rightarrow\mathbb{R}_{\geq0}$ in which the SCFT
sits at the point $r=0$ in $\mathbb{R}_{\geq0}$ where the $S^{3}/\Gamma$
collapses to zero size. We can introduce a defect by wrapping a D3-brane on
the radial direction of $\mathbb{R}_{\geq0}$ as well as a torsional 1-cycle of
$S^{3}/\Gamma$. In this case, we expect the topological operator which acts on
such defects to be given by a D3-brane which wraps the boundary torsional
1-cycle as well as a three-dimensional subspace $M_{3}$\footnote{We take $M_3$ to be connected throughout.} of the 6D spacetime.

The procedure for how to work out the TFT generated by our wrapped D3-brane
follows similar steps to those developed in
\cite{GarciaEtxebarria:2022vzq}. Starting from the
D3-brane worldvolume theory, we have the topological
couplings \cite{Douglas:1995bn, Minasian:1997mm}:%
\begin{equation}
\mathcal{S}_{\text{top}}^{D3}=2\pi i \int_{S} \exp(\mathcal{F}%
_{2})\sqrt{\frac{\widehat{A}(T S)}{\widehat{A}(N S)}%
}\left(  C_{0}+C_{2}+C_{4}\right)  ,
\end{equation}
where here, $\mathcal{F}_{2}=F_{2}-B_{2}$, with $F_{2}$ the $U(1)$ gauge field
strength of the D3-brane, and $B_{2}$ the (pullback of) the NS-NS\ 2-form
potential. Additionally, the $C_{m}$ are the pullbacks of RR potentials onto
the worldvolume of the D3-brane. Due to $SL(2,\mathbb{Z})$ duality covariance, we will label the 2-form curvatures as $F_3:= dB_2$ and $F^D_3:=dC_2$. Expanding out, the relevant couplings for us
are, expressed in differential cohomology (see e.g.,
\cite{Freed:2006ya, Apruzzi:2021nmk}),:\footnote{There is a subtlety here due to the fact that the 5-form
field strength is self-dual. For additional discussion on this point, see
e.g., \cite{Belov:2006jd, Belov:2006xj, Heckman:2017uxe}.}%
\begin{equation}\label{eq:topaction}
\mathcal{S}_{\text{top}}^{D3}=2\pi i \int_{S} \breve{F}_{5}%
+\breve{F}^D_{3}\star\mathcal{\breve{F}}_{2}+\breve{F}_{1}\star\left(  \frac
{1}{2}\mathcal{\breve{F}}_{2}\star\mathcal{\breve{F}}_{2}+\frac{1}{24}%
\breve{e}\right),
\end{equation}
with $\breve{e}$ the Euler class of $S$. Comparing with \cite{GarciaEtxebarria:2022vzq},
it will turn out to be important to also track the term involving $\breve{F}_{1} \star \mathcal{\breve{F}}_{2}\star\mathcal{\breve{F}}_{2}$. On the other hand,
the contribution from the Euler term will play little role in the present analysis.\footnote{It can
play a role in situations where we demand specific Spin / Pin structures for $M_3$.}

One might ask how a term involving $F_1$ arises from purely field theoretic considerations.
Indeed, because there are no continuous marginal parameters in 6D SCFTs \cite{Louis:2015mka, Cordova:2016xhm},
one might be tempted to conclude that no such dependence could be present. Observe, however, that the
6D SCFT admits 4D defects (i.e., codimension two defects) given by D3-brane probes of the local model. The
worldvolume of this D3-brane contains a continuous parameter $\tau$ which is precisely what is also entering
in our generalized symmetry operators.

To proceed further, we need to dimensionally reduce the WZ terms of the D3-brane wrapped
on a torsional cycle $\gamma$ of the extra-dimensional geometry. There is a subtlety here in cases where the $SL(2,\mathbb{Z})$ bundle of an F-theory model
is non-trivial because these duality transformations act on the axio-dilaton as well as the
doublet of 2-form potentials of the IIB background.\footnote{Indeed, even in configurations
where the axio-dilaton is constant, there can still be a non-trivial action on the
2-form potentials of the IIB background.} Consequently, the first case we consider involves
the 6D SCFTs with $\mathcal{N} = (2,0)$ supersymmetry. In these cases the elliptic fibration
is completely trivial, which simplifies the analysis of the D3-brane topological terms.
We next treat the case of single curve non-Higgsable cluster
theories \cite{Morrison:2012np}. In this case, the presence of a non-trivial duality bundle leads us to a
discrete Chern-Simons gauge theory on the generalized symmetry operator, which potentially coupled to background fields. We expect similar considerations to hold
in any background where the axio-dilaton is constant.
In all these cases, we find that 3D defects charged under a suitable 3-form symmetry detect a non-invertible symmetry,
namely the fusion algebra for the symmetry operators contains multiple summands.

The most general situation in which the axio-dilaton is position dependent is, by the same reasoning, expected to also lead to non-invertible symmetries. We anticipate that more general possibilities can arise once we consider topological operators which are also fused with those associated with the 0-form and 1-form symmetries of these 6D SCFTs. These generically arise once we take into account the contributions from flavor 7-branes (see e.g., \cite{Hubner:2022kxr, Cvetic:2022imb, Heckman:2022suy}).

Though we leave the details for future work, it is also clear that we can apply the
same methodology when we compactify a 6D SCFT on a background manifold $Q$ of dimension $l$.
Indeed, all that is required is that we also wrap the topological operator on the relevant cycle (possibly
torsional)\ of $Q$, and again perform the appropriate dimensional reduction.

\subsection{6D $\mathcal{N}=(2,0)$ Theories}\label{ssec:6d2comma0}

As a first class of examples, consider the 6D $\mathcal{N} = (2,0)$ SCFTs as engineered by
type IIB on an ADE singularity $\mathbb{C}^2 / \Gamma$ with $\Gamma$ a finite subgroup of $SU(2)$.
We begin by considering the case $\Gamma$ a cyclic group and then turn to the case of $\Gamma$ non-abelian.

\paragraph{$\Gamma$ Cyclic}
Consider first the case where $\Gamma$ is a cyclic group. The topological field theory of the operator constructed from the D3-brane is then derived by reduction in differential cohomology on the quotient $S^3/\Gamma$. Let us denote the cohomology generators of $S^3/\Gamma$ by $1,u_2,\textnormal{vol}$ in degree 0,2,3 and their lift to differential cohomology by $\breve{1},\breve{u}_2,\textnormal{\u{v}ol}$. We expand as
\begin{equation}\label{eq:expansion}
\begin{aligned}
    \breve{F}_5&=\breve{a}_2 *\textnormal{\u{v}ol}+\breve{a}_3 * \breve{u}_2+\breve{a}_5 * \breve{1}   \\
    \breve{F}_3&= \breve{b}_0 * \textnormal{\u{v}ol}+\breve{b}_1 * \breve{u}_2+\breve{b}_3 * \breve{1} \\
    \breve{F}_2&= \breve{c}_0 * \breve{u}_2+\breve{c}_2 * \breve{1} \\
    \breve{F}_1&= \breve{d}_1 * \breve{1}
    \end{aligned}
\end{equation}
and similarly for $\breve{F}_2^{D},\breve{F}_3^{D}$, where the ``D'' superscript refers to the field strength obtained under an S-duality transformation. The coefficients multiplying $\breve{u}$ are background fields for the discrete symmetries
\begin{equation}\label{eq:sym1}
\begin{aligned}
\breve{a}_3 \leftrightarrow \mathbb{Z}_N^{(2)}\,, \qquad \breve{b}_1 \leftrightarrow \mathbb{Z}_N^{(0)} \,, \qquad \breve{c}_0 \leftrightarrow \mathbb{Z}_N^{(-1)} \,,
\end{aligned}
\end{equation}
while those multiplying $ \breve{1}$ are field strengths for the continuous abelian symmetries
\begin{equation}\label{eq:sym2}
\begin{aligned}
\breve{b}_3 \leftrightarrow U(1)^{(1)} \,, \qquad \breve{c}_2 \leftrightarrow U(1)^{(0)} \,, \qquad \breve{d}_1 \leftrightarrow U(1)^{(-1)}
\end{aligned}
\end{equation}
In the above, the superscripts $(s)$ refer to the corresponding $s$-form symmetry. The expansion along $\textnormal{\u{v}ol}$ is a standard reduction and as $S^3/\Gamma$ has formally infinite volume $\breve{a}_2,\breve{b}_0$ are non-dynamical, measuring fluxes which are absent in the purely geometric background (and therefore vanish). The self-duality of $\breve{F}_5$ implies the vanishing of $\breve{a}_5$. Regarding the axio-dilaton, the curvature of  $\breve{d}_1$ is identified with\footnote{The righthand side is not exact since $\tau$ is not single-valued.} $R(\breve{d}_1)=d(\mathrm{Re}(\tau))\in H_1(M_3,\mathbb{Z})$\footnote{The map $R$ on the differential cohomology group $\breve{H}^p(M_3)$ is part of the short exact sequence $0\rightarrow H^{p-1}(M_3,U(1))\rightarrow \breve{H}^p(M_3)\xrightarrow{R} \Omega^p_\mathbb{Z}(M_3)\rightarrow 0$ where $\Omega^p_\mathbb{Z}(M_3)$ denotes $p$-forms on $M_3$ with integer periods, i.e. where standard $U(1)$ fluxes live. For more details on the basics of differential cohomology geared towards physicists see \cite{Freed:2006ya, Freed:2006yc}, as well as section 2 of the recent paper \cite{Apruzzi:2021nmk}.} and when this class is trivial, then the data contained in the differential cohomology class $\breve{d}_1$ is simply $\tau$. With this in mind, we will also employ a slight redefinition of the $\breve{F}_3$ fields to get rid of the cumbersome $\tau$ factors in the D3 topological action which is $F^{new}_3\equiv \frac{1}{\tau} F^{old}_3$ and $(F^D)^{new}_3\equiv \tau (F^D_3)^{old}$. Consistency with Dirac quantization follows from lifting these fields to M-theory on a torus fibration in the standard duality dictionary, i.e., we interpret the type IIB $SL(2,\mathbb{Z})$ covariant 3-form flux as an M-theory 4-form flux reduced on the elliptic fiber.

We emphasize now that in this case there is no flux non-commutativity contrary to the setup in \cite{GarciaEtxebarria:2019caf}. To see why, consider the Hamiltonian formulation by writing $M_3 = N_2 \times \mathbb{R}_t$. We get the pair of electric and magnetic flux operator valued in the $\text{Tor}\,H^2(N_2 \times \gamma; \mathbb{Z})$ as $\Phi_e(b_0 \star u_2)$ and $\Phi_m(c_0 \star u_2)$. Now on $N_2$, the Poincar\'{e} duals $PD[b_0]$ and $PD[c_0]$ do not intersect for degree reasons, so $\Phi_e(b_0 \star u_2)$ commutes with $\Phi_m(c_0 \star u_2)$ and there are no terms involving co-boundaries giving rise to non-commutativity upon quantization as in \cite{GarciaEtxebarria:2022vzq} that describes a discrete gauge theory in the sense of \cite{Banks:2010zn}.

We insert the expansion \eqref{eq:expansion} into our expression for the topological action to find:
\begin{equation}\label{eq:SD3top}
\begin{aligned}
\mathcal{S}_{\textnormal{top}}^{\textnormal{D3}}=\frac{2\pi i}{N}\int_{M_3} \bigg( a_3 + c_0 \cup b^D_3 + c_0^D\cup b_3+c_2 \cup b^D_1 + c_2^D \cup b_1   \\
-c_0\cup b_3-c^D_0\cup b^D_3-c_2\cup b_1-c^D_2\cup b^D_1\bigg),
 \end{aligned}
\end{equation}
where we have added terms derived from similar expansions for $\breve{F}_2^{D}$ to restore invariance under S-duality. Notice that terms coming from expanding $\breve{F}_3^{D}$ are already present in the $\frac{1}{2}\breve{F}_1\star\mathcal{\breve{F}}_2\star \mathcal{\breve{F}}_2$ term in \eqref{eq:topaction}. The above action simplifies after defining linear combinations given by $b_1'\equiv b^D_1-b_1$, $b_3'\equiv b^D_3-b_3$, $c_0'\equiv c_0-c^D_0$, and $c_2'\equiv c_2-c^D_2$ after which the topological action is just
\begin{equation}\label{eq:SD3top2}
\begin{aligned}
\mathcal{S}_{\textnormal{top}}^{\textnormal{D3}}=\frac{2\pi i}{N}\int_{M_3} \left( a_3 + c'_0 \cup b'_3 +c'_2 \cup b'_1\right).
 \end{aligned}
\end{equation}

Recall that $\breve{F}_2$ and $\breve{F}_2^D$ are worldvolume field strengths on the D3-brane at infinity and therefore $c_0,c_2$ and their dual partners are path-integrated over. The topological operator therefore takes the form:
\begin{equation}
\begin{aligned}
\mathcal{U}(M_3)= \frac{1}{\mathcal{K}}\int Dc'_0 Dc'_2  \exp\left(  \mathcal{S}_{\textnormal{top}}^{\textnormal{D3}}\right)
 \end{aligned}
\end{equation}
where $\mathcal{K}$ is a normalization constant we determine shortly. In our definition of $\mathcal{U}(M_3)$,
we have left implicit the dependence on the torsional 1-cycle of the boundary geometry (to avoid cluttering
notation). At this point, unless otherwise stated, we assume that this torsional 1-cycle is a generator of
$H_1(S^3 / \Gamma, \mathbb{Z})$. The topological operator $\mathcal{U}(M_3)$ is a product of the operator
\begin{equation}
\begin{aligned}
\mathcal{U}_0=\exp\left( \frac{2 \pi i}{N} \int_{M_3} a_3 \right)\,,
\end{aligned}
\end{equation}
which is the standard flux operator for surface defects of the SCFT, and
\begin{equation}
\begin{aligned}
\mathcal{U}_1&=\frac{1}{|H_1(M_3, \mathbb{Z}_N)|}\int Dc'_2 \exp\left( \frac{2 \pi i}{N} \int_{M_3} c'_2\cup b'_1 \right)\,, \\
\mathcal{U}_3&=\frac{1}{N}\int Dc'_0 \exp\left( \frac{2 \pi i}{N} \int_{M_3} c'_0\cup b'_3 \right)\,.
\end{aligned}
\end{equation}
So altogether we have
\begin{equation}
    \mathcal{U}(M_3) = \mathcal{U}_0 \mathcal{U}_1 \mathcal{U}_3 \,.
\end{equation}
which sets the normalization constant $\mathcal{K}$. When the $C_2$ and $B_2$ backgrounds are turned off we have, $\mathcal{U}(M_3) = \mathcal{U}_0$.

Let us now study the fusion algebra. Note that all operators except $\mathcal{U}_0$ are condensation operators since they specify a 3-gauging of a $U(1)^{(3)}$ or $\mathbb{Z}^{(4)}_N$ symmetry along the $M_3$ worldvolume. Moreover, we show that these operators satisfy the fusion algebra of projections
\begin{equation}
    \mathcal{U}_i \mathcal{U}_i = \mathcal{U}_i\,, \; \; \; \; \textnormal{($i=1,3$)},
\end{equation}
and so formally speaking are non-invertible. That being said, they are invertible when
restricted to their image where they equate to the identity operator. This follows
for instance for $\mathcal{U}_1$ by the manipulations
\begin{equation}
\begin{aligned}
    \mathcal{U}_1&=\frac{1}{|H_1(M_3, \mathbb{Z}_N)|}\int Dc'_2 \exp\left( \frac{2 \pi i}{N} \int_{M_3} c'_2\cup b'_1 \right) \\
    &=\frac{1}{|H_1(M_3, \mathbb{Z}_N)|}\sum_{\ell \in H_1(M_3, \mathbb{Z}_N)} \exp\left( \frac{2 \pi i}{N} \int_\ell b'_1 \right)\\
     &=\frac{1}{|H_1(M_3, \mathbb{Z}_N)|}\prod_{\ell'} \left(\sum_{k=0}^{N-1} \exp\left( \frac{2 \pi i k}{N} \int_{\ell'} b'_1 \right)\right)
    \end{aligned}
\end{equation}
together with the integrality of the periods of $b'_1$. Here $\{\ell'\}$ are a generating set for the lattice $H_1(M_3, \mathbb{Z}_N)$. So whenever such periods are non-vanishing we have a vanishing sum of roots of unity.
From this we also see that $\mathcal{U}_i= \mathcal{U}_i^\dagger$ for $i\neq 0$ follows from relabeling $k\rightarrow -k$. The normalization is now explicitly $\mathcal{K}=N|H_1(M_3, \mathbb{Z}_N)|$. On the other hand the operator $\mathcal{U}_0$
displays a cyclic fusion ring
\begin{equation}
\begin{aligned}
\mathcal{U}_0 \mathcal{U}_0^\dagger =1\,, \qquad \mathcal{U}_0^n =\exp\left(\frac{2 \pi i n}{N} \int_{M_3} a_3 \right).
\end{aligned}
\end{equation}

Concerning the operators charged under $\mathcal{U}(M_3)$, these include the surface operators of the defect group $\mathbb{D}$ constructed from D3-branes wrapped on relative 2-cycles of the F-theory base $\mathcal{B}$.
The operators $\mathcal{U}_1, \mathcal{U}_3$ do not act on elements of $\mathbb{D}$ since they carry no charge under the symmetries of \eqref{eq:sym1} and \eqref{eq:sym2} other than $\mathbb{Z}_N^{(2)}$. Therefore the restriction of $\mathcal{U}(M_3)$ on $\mathbb{D}$ is given by $\mathcal{U}_0$, namely the standard flux operator. However, as mentioned at the end of Section \ref{sec:branesandgensym}, $\mathcal{U}(M_3)$ can act on operators with spacetime dimension other than $2$ as well. The $\mathcal{U}_1$ piece acts on local operators of the 6D SCFT that originate from $D1$ and $F1$ strings wrapping a torsional 1-cycle in the boundary $S^3/\Gamma$ times the radial direction of $\mathbb{C}^2/\Gamma$, while the $\mathcal{U}_3$ piece acts on line operators that wrap a point in $S^3/\Gamma$ times the radial direction. The actions of $\mathcal{U}_{1}$ and $\mathcal{U}_{3}$ on these operators is almost trivial: it multiplies by zero on any operators with non-zero charge under the symmetry groups $\mathbb{Z}^{(0)}_N$ and $U(1)^{(1)}$ respectively.

\paragraph{$\Gamma$ Non-Abelian}
Consider next the case where $\Gamma$ is non-abelian. As far as the defect group is concerned, the relevant data is captured by the abelianization $\mathrm{Ab}[\Gamma]$. Returning to the entries of Table \ref{tab:defectgrps}, we see that in nearly all cases, we again have a single cyclic group factor so the analysis proceeds much as we already presented. On the other hand, for some $D$-type subgroups,
$\mathrm{Ab}[\Gamma]$ has two cyclic group factors. For this reason, we now focus on this case.

Proceeding more generally, when we insert the above expansion (\ref{eq:expansion}) into (\ref{eq:topaction}),
the overall coefficient we obtain in the exponential is given by the canonical link pairing in first homology:
\begin{equation}
    L_\Gamma: \; H_1(S^3/\Gamma)\times H_1(S^3/\Gamma)\rightarrow \mathbb{Q}/\mathbb{Z}.
\end{equation}
This is because given $\breve{t}^i_2$ such that $I(\breve{t}^i_2)=t^i_2\in H^2(S^3/\Gamma,\mathbb{Z})\simeq \mathbb{Z}_{n_i}\times \mathbb{Z}_{n_j}$ we have that
\begin{equation}
\int_{S^3/\Gamma}\breve{t}^i_2 * \breve{t}^j_2 =L^{ij}_\Gamma.
\end{equation}
Table \ref{tab:defectgrps} gives the explicit linking pairing for all $\Gamma$ a finite subgroup of $SU(2)$.\footnote{In the more general case where $\Gamma$ is a $D_{p+q,q}$ subgroup of $U(2)$ and $\mathrm{Ab}[\Gamma]$ has two cyclic group factors, determining the linking pairing is somewhat dependent on the divisibility properties of $p$, $q$ and $p+q$. The linking pairing was worked out on a case by case basis in some examples in reference \cite{GarciaEtxebarria:2019caf}. There, one can see that $L^{ij}_\Gamma$ can be recast as an intersection pairing of certain non-compact 2-cycles in a blow-up of $\mathbb{C}^2/\Gamma$. It is tempting to speculate that one can use a quiver-based method to directly read off this data, much as in \cite{DelZotto:2022fnw}.}

Let us turn next to the TFT obtained from wrapping a D3-brane on a torsional cycle of $S^3 / \Gamma$. When $\breve{H}^{2}(S^3 / \Gamma)$ has more than one generator, the previously considered $\breve{a}_3$, $\breve{b}_1$, and $\breve{c}_0$ each pick up an index. In determining the spectrum of topological operators, it is enough to consider D3-branes wrapping $\gamma = \nu^{1} \gamma_{1} + \nu^{2} \gamma_{2}$, with $\gamma_{i}$ a primitive generators of $H_{1}(S^3 / \Gamma)$. The action is now (reverting back to the original duality basis for clarity):
\begin{equation}
    \begin{aligned}
\mathcal{S}_{\textnormal{top}}^{\textnormal{D3}}=2\pi \sqrt{-1} \nu^{i} (L_\Gamma)_{ij} \underset{M_3}{\int} \bigg( a^j_3 + c^j_0 \cup b^D_3 + (c_0^D)^j\cup b_3+c_2 \cup (b_1^D)^j  + c_2^D \cup b^j_1  \\
-c^j_0 \cup b_3 - (c_0^D)^j\cup b^D_3-c_2 \cup b^j_1 - c_2^D \cup (b_1^D)^j
\bigg).
 \end{aligned}
\end{equation}
Just as in the case of $\Gamma = \mathbb{Z}_N$, we observe that the fusion rules for these topological operators produce an invertible symmetry when the background $C_2$ and $B_2$ fields are switched off.

\subsection{6D NHC Theories} \label{sec:6d10}

We now turn to rank one 6D $\mathcal{N}  = (1,0)$ theories in which the axio-dilaton is constant
but the duality bundle of the F-theory model is still non-trivial. In particular, we consider the
case of the single curve non-Higgsable clusters (NHCs) of reference \cite{Morrison:2012np}
in which the base of the F-theory
model supports a curve of self-intersection $-n$ for $n=3,4,6,8,12$. These models
can be written as $(\mathbb{C}^2 \times T^2) / \mathbb{Z}_n$, where the action on the $\mathbb{C}^2$
base is by a common $n^{th}$ primitive root of unity \cite{Witten:1996qb, Heckman:2013pva}.
On the tensor branch, these theories are characterized by a 6D gauge theory coupled to a tensor multiplet with charge
prescribed by the self-intersection number. With notation as in \cite{Heckman:2013pva}, we have:
\begin{equation}
    \overset{\mathfrak{su}(3)}{3}, \;\;  \overset{\mathfrak{so}(8)}{4}, \;\;  \overset{\mathfrak{e}_6}{6}, \;\;  \overset{\mathfrak{e}_7}{8}, \;\;  \overset{\mathfrak{e}_8}{12}
\end{equation}
where $\overset{\mathfrak{g}}{n}$ refers to a $(-n)$-curve with a ADE singularity of type $\mathfrak{g}$ wrapping it.
These can all be presented as F-theory backgrounds $(\mathbb{C}^2\times T^2)/\mathbb{Z}_n$ (see \cite{Witten:1996qb, Heckman:2013pva}) where the quotient is defined by the group action:
\begin{equation}
    (z_1,z_2,z_3)\rightarrow (\zeta_n z_1,\zeta_n z_2, \zeta^{-2}_n w )
\end{equation}
where $w$ is the torus-fiber coordinate. In the nomenclature of table \ref{tab:defectgrps}, $\Gamma=\mathbb{Z}_n{(1)}$ (i.e., $p = n$ and $q = 1$), and thus the link pairing is $L_{\Gamma}=1/n$.

To build $\mathcal{U}(M_3)$, we again wrap a D3 on $M_3\times \gamma$ where $\gamma$ is a generating 1-cycle with boundary homology class $\gamma \in H_1( S^3/\Gamma)$, but now there is a non-trivial $SL(2, \mathbb{Z})$ monodromy for $n=3,\; 4, \; 6, \; 8$ and $12$. This clearly modifies the expansion of $\breve{F}_3$, $\breve{F}_2$, and their duals in such a way that one must generally consider the vectors $(F_{2},F^D_{2}), (F_{3},F^D_{3})$ modulo some relations as well-defined objects rather than the individual components. More precisely, we need to expand these fields in cohomology with the twisted coefficient module
\begin{equation}\label{eq:twistcoef}
    (\mathbb{Z} \oplus \mathbb{Z})_\rho
\end{equation}
where $\rho$ is the $SL(2, \mathbb{Z})$ monodromy of order $k$ when going around $\gamma$ as given in table \ref{tab:TopData}.

\begin{table}
\begin{center}
\renewcommand{\arraystretch}{1.25}
\begin{tabular}{|| c | c | c | c | c | c ||}
 \hline
$n$ & \textnormal{Kodaira Type} & Monodromy $\rho$ & $k= \mathrm{ord}(\rho)$ & Tor\,$H_1(T_3)$ &  $L_\gamma^t$ \\ [0.5ex]
 \hline\hline
 3 & $IV$ &  $\left(\begin{array}{cc}  0 & 1 \\ -1 & -1 \end{array}\right)$ & 3 &  $\mathbb{Z}_3$ & 1/3  \\
 \hline
 4 & $I_0^*$ & $\left(\begin{array}{cc}  -1 & 0 \\ 0 & -1 \end{array}\right)$ & 2  &  $\mathbb{Z}_2 \oplus \mathbb{Z}_2$ & $\left(\begin{array}{cc}  0 & 1/2 \\ 1/2 & 0 \end{array}\right)$ \\
 \hline
 6 & $IV^*$ &  $\left(\begin{array}{cc}  -1 & -1 \\ 1 & 0 \end{array}\right)$ & 3  &  $\mathbb{Z}_3$ & 2/3 \\
 \hline
 8 & $III^*$ & $\left(\begin{array}{cc}  0 & -1 \\ 1 & 0 \end{array}\right)$ & 4 &  $\mathbb{Z}_2$ & 1/2  \\
 \hline
 12 & $II^*$ & $\left(\begin{array}{cc}  0 & -1 \\ 1 & 1 \end{array}\right)$ & 6 &  $0$ & $0$  \\
 \hline
\end{tabular}
\end{center}
\caption{Topological data for asymptotic geometries of 6D $\mathcal{N}=(1,0)$ SCFTs. The F-theory geometry consists of an elliptic fibration over a Lens space base containing a torsional 1-cycle $\gamma$. The torus fibration restricted to $\gamma$ gives a three-manifold $T_3$, a torus bundle more precisely, whose torsion and linking forms $L^t_\gamma$ we list, see \cite{Cvetic:2021sxm}. D3 branes wrapped on $\gamma$ map under M-/F-theory duality to M5-branes wrapped on $T_3$. }
\label{tab:TopData}
\end{table}

We begin by computing the twisted cohomology of the boundary of the base space $S^3/\Gamma = \partial(\mathbb{C}^2/\Gamma)$. Via an identical computation\footnote{Let $A = \mathbb{Z}^2$ be a $\mathbb{Z}_n$ module, then $H^*(S^{2r-1}/\mathbb{Z}_n, A)$ is computed by taking the cohomology of the cochain complex:
\begin{equation}
\mathbb{Z}^2 \xrightarrow[]{\,1-\rho\,} \mathbb{Z}^2 \xrightarrow[]{\,1+\rho+\cdots+\rho^{n-1}\,}  \mathbb{Z}^2 \xrightarrow[]{\,1-\rho\,} \cdots \xrightarrow[]{\,1+\rho+\cdots+\rho^{n-1}\,} \mathbb{Z}^2 \xrightarrow[]{\,1-\rho\,} \mathbb{Z}^2\,.
\end{equation}
} to that given in section 3.2 of \cite{Aharony:2016kai}, we get
\begin{align}
    H^*(S^3/\mathbb{Z}_n; (\mathbb{Z} \oplus \mathbb{Z})_\rho) = \{0, G_k, 0, G_k\}
\end{align}
where
\begin{equation}\label{eq:discgrps}
    G_k = \left\{
   \begin{array}{lll}
    \mathbb{Z}_2 \oplus \mathbb{Z}_2 &\quad k=2  &\quad (n = 4) \\
    \mathbb{Z}_3 &\quad k=3  &\quad  (n = 3,6) \\
    \mathbb{Z}_2 &\quad k=4 &\quad  (n = 8) \\
        1 &\quad k=6 &\quad  (n = 12)
    \end{array}
    \right.
\end{equation}

Now, $\breve{F}_5$ associated with D3 branes should be reduced on untwisted differential cocycles $\breve{u}_i \in \breve{H}^i(S^3/\Gamma; \mathbb{Z})$:
\begin{equation}
        \breve{F}_5 = \breve{a}_2 \star \textnormal{\u{v}ol} + \breve{a}_3 \star \breve{u}_2 + \breve{a}_5 \star \breve{1}, \\
\end{equation}
whereas $\breve{F}_3, \breve{F}_3^{(D)}$ associated with D1 and F1 strings and the worldvolume  $\breve{F}_2, \breve{F}_2^{(D)}$ should be reduced on twisted differential cocycles $\breve{t}_i \in \breve{H}^i(S^3/\Gamma; (\mathbb{Z} \oplus \mathbb{Z})_\rho)$. The reduction goes as follows: when $k\neq 2$ (here $\rho$ in the superscript stands for self-dual operators that are compatible with the $\rho$ twisting):
\begin{align}
\label{eq:Monodromy}
    \begin{split}
     \breve{F}_3^{\rho} = \left(\begin{array}{c} \breve{F}_3 \\ \breve{F}_3^{(D)} \end{array}\right)/\mathrm{Im}(\rho - 1) &= \breve{b}_2^{\rho} \star \breve{t}_1 + \breve{b}_0^{\rho} \star \breve{t}_3  \\
    \breve{F}_2^{\rho} = \left(\begin{array}{c} \breve{F}_2 \\ \breve{F}_2^{(D)} \end{array}\right)/\mathrm{Im}(\rho - 1) &= \breve{c}_1^{\rho} \star \breve{t}_1\,.
    \end{split}
\end{align}
Similar expansions hold for $ \breve{b}_0^{\rho}$ and $\breve{b}_2^{\rho}$. The notation of the lefthand side denotes the reduction of $(\breve{F}_2, \breve{F}_2^{(D)})$ modulo $SL(2, \mathbb{Z})$ monodromy. The fields $c_{n}$ and $b_{m}$ are discrete $G_k$ valued $n$-cocycles and $m$-cocycles where $G_k=\mathbb{Z}_2,\mathbb{Z}_3$ as in \eqref{eq:discgrps} respectively.

When $k=2$ we have the decomposition
\begin{equation}
H^i(S^3/\Gamma; (\mathbb{Z} \oplus \mathbb{Z})_\rho)
=H^i(S^3/\Gamma; (\mathbb{Z})_\rho)^{(e)}\oplus H^i(S^3/\Gamma; (\mathbb{Z})_\rho)^{(m)}\end{equation}
with $\rho=-1$ for each coefficient ring. Consequently the expansion is then
\begin{equation}
    \breve{F}_2 =\breve{c}_1 \star \breve{t}_{(e),1} \,, \qquad
    \breve{F}_2^{(D)}= \breve{c}_1^D \star  \breve{t}_{(m),1}
\end{equation}
where $I(\breve{t}_{(e,m),1})=t_{(e,m),1}$ generates $H^1(S^3/\Gamma; (\mathbb{Z})_\rho)^{(e,m)}$. Similar expansions hold for $ \breve{b}_0$ and $\breve{b}_2$. In this case all fields are discrete $\mathbb{Z}_2$ co-cycles with the degree as indicated by their index.

When reducing the $\breve{F}_5$ term, the non-zero term comes from
\begin{equation}
     \int_{S^3/\Gamma} \breve{u}_2 \star \breve{u}_2 \equiv L_\Gamma(\breve{u}_2)
\end{equation}
on $S^3/\Gamma$. This gives the contribution to the action of
\begin{equation}
    \exp \left( 2\pi i L_\Gamma(\breve{u}_2) \int_{M_3} a_3  \right).
\end{equation}

For $\breve{F}_3, \breve{F}_3^{(D)}, \breve{F}_2, \breve{F}_2^{(D)}$, on the other hand, the non-zero terms can be evaluated from the pairing of the twisted cohomology classes on $\gamma \in H_1(S^3/\Gamma; \mathbb{Z})$:
\begin{equation}\begin{aligned}\label{eq:Pairing}
 k=3,4\,:\qquad &\int_{\gamma} \breve{t}_1 \star \breve{t}_1 \equiv L_\gamma^{t}(\breve{t}_1)\\ k=2\,:\qquad &\int_{\gamma} \breve{t}_{(e),1} \star \breve{t}_{(m),1} \equiv L_\gamma^{t}(\breve{t}_{(e),1},\breve{t}_{(m),1})
\end{aligned}\end{equation}
which we  both denote by $L_\gamma^t$ whenever the context is clear. The self-pairing of $\breve{t}_{(e,m),1}$ vanishes as we shortly argue.

The pairing between twisted classes in differential cohomology generalizing torsional linking are computed using the methods in reference \cite{Bott1982DifferentialFI}. M-/F-theory duality gives a natural relation of such pairings to linking forms in ordinary singular homology. We now explain this relation as we perform our computations from the latter perspective.

To frame the discussion we introduce the torus bundle three-manifold $T_3$ as the restriction of the S-duality torus bundle to the 1-cycle $\gamma$ wrapped by the D3 brane. As all three-manifolds its homology groups are fully determined by $H_1(T_3;\mathbb{Z})$ which is computed by application of the Mayer-Vietoris sequence to
\begin{equation}
    H_1(T_3)=\mathbb{Z}\oplus \textnormal{coker}\,(\rho-1)
\end{equation}
where $\rho$ is the $SL(2,\mathbb{Z})$ monodromy matrix acting on 1-cycles upon traversing $\gamma$. The torsional subgroups are listed in table \ref{tab:TopData}. The linking form on $T_3$ follows from similar considerations \cite{Cvetic:2021sxm}. Let us denote the torsional generators of $\textnormal{Tor}\,H^2(T_3;\mathbb{Z})$ by $t_2$ which by Poincar\'e duality and the universal coefficient theorem is dual to the generators $\ell_1$ of $\textnormal{Tor}\,H_1(T_3;\mathbb{Z})$. When the monodromy matrix is of type $I_0^*$ $(k=2)$ both $t_2$ and $\ell_1$ are further indexed by $(e,m)$ distinguishing the factors in $\textnormal{Tor}\,H_1(T_3;\mathbb{Z})\cong \mathbb{Z}_2 \oplus \mathbb{Z}_2$ for that case.

Now note that M-/F-theory duality maps a D3 brane wrapped on $\gamma\times M_3$ to an M5-brane wrapped on $T_3\times M_3$ where $M_3$ is the space-time submanifold supporting the topological operator. The Wess-Zumino-Witten term of the M5-brane contains the term \cite{Aharony:1996wp,  Bandos:1997ui}
\begin{equation}\label{eq:WZWM5}\begin{aligned}   \mathcal{S}_{\text{top}}^{M5} \supset  2\pi i \int_{M_3\times T_3}  \breve{F}_7 + \frac{1}{2} \breve{F}_3 \star \breve{F}_4
\end{aligned}\end{equation}
where $\breve{F}_3$ is the anti-self-dual 3-form field strength (of the anti-chiral 2-form field on the M-theory worldvolume), and $\breve{F}_7$ is the pullback of the magnetic dual 7-form field strength.
We can therefore equivalently compute the topological field theory on $M_3$ starting from the action \eqref{eq:WZWM5}. This approach however expresses the coefficient of the topological theory via geometric data of $T_3$ and avoids $SL(2,\mathbb{Z})$ twisted cohomology classes. We therefore conjecture that the pairing \eqref{eq:Pairing} is geometrized to a link pairing on the torus bundle
\begin{equation}\label{eq:Conjecture}\begin{aligned}
   k=3,4\,:\qquad &L_\gamma^{t}(\breve{t}_1)= \int_\gamma \breve{t}_1\star\breve{t}_1=\int_{T_3} \breve{t}_2\star\breve{t}_2\\
      k=2\,:\qquad &L_\gamma^{t}(\breve{t}_{(e),1},\breve{t}_{(m),1})= \int_\gamma \breve{t}_{(e),1}\star\breve{t}_{(m),1}=\int_{T_3} \breve{t}_{(e),2}\star\breve{t}_{(m),2}=\frac{1}{2}
   \end{aligned}
\end{equation}
where the righthand side is computed by the linking pairing given in table \ref{tab:TopData}. We will have more to say on the M-theory perspective in section \ref{sec:Mtheory}. Evidence for the identity \eqref{eq:Conjecture} is already given in \cite{GarciaEtxebarria:2022vzq} which considers a setup with $-1\in SL(2;\mathbb{Z})$ monodromy along $\gamma$ and where the case $k=2$ in \eqref{eq:Conjecture} was found to hold.

Before writing down the full topological action of our D3-brane, we must first comment on the expected non-commutativity of flux operators in this scenario. Due to the presence of a non-trivial duality bundle, there is a mixing between the electric and magnetic dynamical two-form curvatures on the D3 worldvolume gauge theory, so considering the on-shell relation $F^D_2=*F_2$ (see for instance \cite{Polchinski:2014mva}) we must quantize these fields as a self-dual Maxwell theory.\footnote{Note that our worldvolume theory is Euclidean.} This is especially clear in the M5-brane picture where the flux quantization is already that of anti-self-dual fields and the torsion homology of the $T^2$-bundle precisely descends to the torsion in the twisted homology that the D3-brane wraps. Depending on the value of $k$, it is occasionally possible to have a canonical splitting of electric and magnetic fluxes on the D3.

 Now expanding on the treatment of \cite{GarciaEtxebarria:2022vzq} to examine the non-commutativity of fluxes on the D3 worldvolume in our cases, we first assume that $M_3=N_2\times \mathbb{R}_t$ to employ a Hamiltonian formalism. The Hilbert space associated to the spatial manifold $N_2 \times \gamma$ of the D3-brane worldvolume will then be a representation a Heisenberg algebra, the details of which depend on the value of $k$. The Heisenberg algebra is generated by non-commuting electric and magnetic flux operators $\Phi_e,\Phi_m$ respectively detecting fluxes through torsional cycles. The cases are:\footnote{We thank I. Garcia Etxebarria for a question which prompted this clarification. See also \cite{GarciaHosseini} for a related discussion.}
\begin{itemize}
    \item For $k=2$, we have a pair of non-commuting electric and magnetic fluxes associated with $\mathrm{Tor}\, H^2(N_2 \times \gamma)$:
    \begin{equation}
        \Phi_e( \breve{c}_1 \star \breve{t}_{(e),1}) \Phi_m(\breve{c}_1^{(D)} \star \breve{t}_{(m),1}) = \exp\left( 2\pi iL_\gamma^{t}(\breve{t}_{(e),1},\breve{t}_{(m),1}) \right) \Phi_m(\breve{c}_1^{(D)} \star \breve{t}_{(m),1})  \Phi_e( \breve{c}_1 \star \breve{t}_{(e),1})
    \end{equation}
    we thus get a $\mathbb{Z}_2$ gauge theory as in \cite{Freed:2006ya}, with the action given by
    \begin{equation}
    \mathcal{S}_{\mathbb{Z}_2} =  \pi i \int_{M_3}c_1 \cup \delta c_1^{(D)}
    \end{equation}
    where $(c_1, c_1^{(D)})$ are a pair of discrete gauge fields which together are valued in $G_2=\mathbb{Z}_2 \times \mathbb{Z}_2$. In other words, $c_1$ and $c_1^{(D)}$ are each separately $\mathbb{Z}_2$ valued discrete gauge fields normalized such that $\int c_1=\ell \; \mathrm{mod}\; 2$.

    \item For $k=3, 4,6$, we have a pair of non-commuting self-dual fluxes associated
    with $\mathrm{Tor}\,H^2(N_2 \times \gamma)$.
    Now, extra care has to be taken since the electric and magnetic field has to be the same \cite{GarciaEtxebarria:2019caf}. For $x, y \in H^{1}(\gamma; (\mathbb{Z} \oplus \mathbb{Z})_\rho)$
    \begin{equation}
        \Phi_a(\breve{c}_1^{\rho} \star \breve{x}) \Phi_b(\breve{c}_1^{\rho} \star \breve{y}) = \exp\left( 2\pi i L_\gamma^{t}(\breve{x}, \breve{y}) \right)   \Phi_b(\breve{c}_1^{\rho} \star \breve{y}) \Phi_a(\breve{c}_1^{\rho} \star \breve{x})
    \end{equation}
    here $L^t_\gamma(x, y)$ is the bilinear form of the twisted linking pairing. 
    Thus, we need to include a discrete CS gauge theory of the form
    \begin{equation}
        \mathcal{S} = 2 \pi i L^t_{\gamma}(\breve{t}_1) \int_{M_3} c_1^\rho \cup \delta c_1^\rho,
    \end{equation}
    where $c^\rho_1$ is a discrete gauge field valued in $G_k$ as in (\ref{eq:discgrps}) normalized such that $\int c^\rho_1=\ell \; \mathrm{mod}\; \mathrm{2\;  or\;  3}$.
\end{itemize}

Furthermore, the topological actions generating the link pairing above produce an additional term in the effective action of the D3 brane reduced on the twisted 1-cycle, the middle term(s) in both lines of \eqref{eq:SD3toptwist}, because the effective action must be a functional of the gauge invariant combination $\delta c^\rho_1-b^\rho_2$ where $b^\rho_2$ is defined in the first line of equation \eqref{eq:Monodromy}.\footnote{This follows from the fact that the standard D3 topological action must be a functional of the gauge invariant combination $F-B_2$ and its S-dual completion $F^D-C_2$.}

To summarize then, we get the action of the topological operator (where again we keep the dependence on the torsional 1-cycle implicit, include a normalization factor $\mathcal{K}^{-1}$, and leave the cup products implicit) $\mathcal{U}(M_3) =\mathcal{K}^{-1} \exp(\mathcal{S}_{\textnormal{top}}^{\textnormal{D3}} )$ as:
\begin{equation} \label{eq:SD3toptwist}
\begin{aligned}
    k=2 &: \qquad  \mathcal{S}_{\textnormal{top}}^{\textnormal{D3}} =
    2\pi i\int_{M_3} \bigg( -\tfrac{1}{4} a_3 - \tfrac{1}{2}b^{(D)}_2 c_1 - \tfrac{1}{2} b_2 c_1^{(D)} + \tfrac{1}{2} c_1 \delta c_1^{(D)} \bigg)
    \\
    k= 3, 4, 6 &:   \qquad \mathcal{S}_{\textnormal{top}}^{\textnormal{D3}} = 2\pi i\int_{M_3} \bigg( L_\Gamma a_3 - L_\gamma^t  b_2^{\rho} c_1^{\rho} + L_{\gamma}^t c_1^{\rho} \delta c_1^{\rho} \bigg)
\end{aligned}
\end{equation}
where for $k = 3, 4, 6$, the path integral is written as $\mathcal{K}^{-1}\int Dc_1^{\rho}  \exp\left(  \mathcal{S}_{\textnormal{top}}^{\textnormal{D3}}\right)$ with the implicit understanding that a delta--function relating $c_1, c_1^{(D)}$ has been inserted to gauge fix the monodromy relations of line \eqref{eq:Monodromy}. For the values of both $L_\Gamma=L_\Gamma(\breve{u}_2) = \frac{1}{n}$ and $L_{\gamma}^t=L_\gamma^t(\breve{t}_1)$ see table \ref{tab:TopData}. Also, in this subsection the normalization factor will be $\mathcal{K}=(|H_2(M_3,G_k)|)^{1/2}$ for reasons that will be clear in what follows.


Notice that due to the coupling to the various $b_2$ fields in line \eqref{eq:SD3toptwist}, we see again, just as in subsection \ref{ssec:6d2comma0}, that the symmetry operator $\mathcal{U}(M_3)$ acts on more than just dimension-2 operators in the defect group $\mathbb{D}$. The $b_2$'s are background fields for a discrete $G_k^{(1)}$-symmetry and are sourced by defects constructed from $(p,q)$-5-branes wrapping homology classes\footnote{Said differently, a 2-cycle with $(p,q)$-charge of the brane is being measured modulo $\textnormal{Im}\,(\rho-1)$.} in $H_2(S^3/\Gamma, (\mathbb{Z}\oplus \mathbb{Z})_\rho)=\mathrm{Hom}(G_k,U(1))$ times the radial direction of $\mathbb{C}^2/\Gamma$.

 Notice that due to the presence of terms like $c\delta c$ in \eqref{eq:SD3toptwist}, we see that even if we ignore terms involving $b_2$ fields (and hence the effect of $\mathcal{U}(M_3)$ on the line operators mentioned in the previous paragraph) our 2-form symmetry operators are tensored with discrete topological gauge theories, namely some level-N Dijkgraaf-Witten theories with gauge group $G_k$, $\mathcal{T}^{(N,G_k)}$. The levels of these gauge theories are classified by by $H^4(G_k,\mathbb{Z})$ \cite{Dijkgraaf:1989pz}, and the possible levels relevant to single node NHCs are $H^4(\mathbb{Z}_N,\mathbb{Z})=\mathbb{Z}_N$ and $H^4(\mathbb{Z}^2_2,\mathbb{Z})=\mathbb{Z}^3_2$. Our action (\ref{eq:SD3toptwist}) thus predicts the levels of these discrete gauge theories living on the defect group symmetry operators and notice that the operator fusion simply adds together the cyclically defined Chern-Simons levels, so this confirms that that $\mathcal{U}(M_3)$ appears to be an invertible operator when linking with operators with trivial charge under the $G^{(1)}_k$ symmetry.

\textbf{Fusion Rules}
We now turn to the fusion rules for our symmetry operators, namely we
compute $\mathcal{U}(M_3)\times \mathcal{U}^\dagger(M_3)$. We find that
in the presence of the $G^{(1)}_k$ background field $b_2$ that the fusion rule typically
contains multiple summands, i.e., the hallmark of a non-invertible symmetry.\footnote{The non-invertible fusion is in fact not very surprising considering that the terms in the actions of (\ref{eq:SD3toptwist}) not involving $a_3$ are
a discrete analog of the 3D actions one would write for the standard fractional Quantum Hall effect (FQHE). This is similar to what was found in the analysis of ABJ anomalies in references \cite{Choi:2022jqy,Cordova:2022ieu}.} This
is detected by 3D defects sourcing such backgrounds and linked by surface operators which in turn are produced in the fusion.
\begin{itemize}
\item For the simplest case of $k = 6$, the symmetry operator associated with a generator of $H_1(S^3 / \mathbb{Z}_{12}, \mathbb{Z})$ takes the form
\begin{equation}
    \mathcal{U}(M_3) = \exp \left( \frac{2\pi i}{12}  \int_{M_3} a_3 \right)
\end{equation}
which simply reproduces the $\mathbb{Z}_{12}$ defect group of a $(-12)$-curve NHC.
\item For $k = 2$ (i.e. the $-4$ NHC theory), we can decompose $\mathcal{U}(M_3) = \mathcal{U}_0(M_3) \mathcal{U}_1(M_3)$. $\mathcal{U}_0(M_3) = \left( \frac{2\pi i}{4}  \int_{M_3} \breve{a}_3 \right)$ generates a $\mathbb{Z}_4$ defect group, whereas $\mathcal{U}_1(M_3) $ is non-invertible when $b_2$ or $b^D_2$ is turned on. Physically, this means
3D defects which are charged under the 1-form symmetry (with background field $b_2$) will detect
this non-invertible fusion rule.

Because our topological action exactly matches that of equation (3.14) of \cite{GarciaEtxebarria:2022vzq} up to an overall irrelevant minus sign, we can borrow the result to state
\begin{equation}\label{eq:kequal2condensation}
 \mathcal{U}_1(M_3)\times  \mathcal{U}_1(M_3)=\frac{1}{|H_2(M_3,\mathbb{Z}_2)|^2}\sum_{\sigma, \sigma'\in H_2(M_3,\mathbb{Z}_2)} \exp{\bigg(\pi i \int_\sigma b_2\bigg)}\cdot \exp{\bigg(\pi i \int_{\sigma'} b^{(D)}_2\bigg)}
\end{equation}
where $\sigma$ and $\sigma'$ are generators of $H_2(M_3, \mathbb{Z}_2)$. Notice that these exponents are symmetry operators for a discrete $\mathbb{Z}_2\times \mathbb{Z}_2$ 1-form symmetry in the 6D SCFT. In the language of \cite{Roumpedakis:2022aik} this sum over symmetry operators restricted to lie in $M_3$ means that this is a 3-gauging of the $\mathbb{Z}_2\times \mathbb{Z}_2$ 3-form symmetry along $M_3$. This is commonly known as a condensation operator.


\item For $k = 3,4$, we can similarly decompose $\mathcal{U}(M_3) = \mathcal{U}_0(M_3) \mathcal{U}_1(M_3)$ where the invertible piece $\mathcal{U}_0(M_3) = \left( \frac{2\pi i}{n}  \int_{M_3} \breve{a}_3 \right)$ produces a $\mathbb{Z}_6$ and $\mathbb{Z}_8$ algebra respectively.\footnote{Recall that $n=6$ for $k=3$ and $n=8$ for $k=4$.} $\mathcal{U}_1(M_3)$ is again non-invertible which we can see from the fact that the fusion product $\mathcal{U}_1(M_3)\times \mathcal{U}^\dagger_1(M_3)\neq \mathbf{1}$ (for $k=2$, $\mathcal{U}_1(M_3)=\mathcal{U}^\dagger_1(M_3)$).
Calculating the total fusion (leaving the correct normalization until the final result),
\begin{align}
    & \mathcal{U}(M_3) \times \mathcal{U}^\dagger(M_3) = \int Dc_1^\rho D c_1^{\prime \rho}  \exp\bigg( 2\pi i L^t_\gamma \int_{M_3} \big( c_1^{\rho} \cup\delta c_1^{\rho} - c_1^{\prime\rho} \cup \delta c_1^{\prime\rho}+b^\rho_2\cup(c^{\prime\rho}_1-c_1^{\rho}) \big) \bigg)
\end{align}
and after substituting $\hat{c}^\rho_1:=c^{\prime\rho}_1-c^{\rho}_1$ and integrating a term by parts we find
\begin{align}\label{eq:uudaggerk34}
   &\mathcal{U}(M_3)\times \mathcal{U}^\dagger(M_3) =\int Dc_1^{\prime \rho} D \hat{c}_1^{ \rho}  \exp\bigg( 2\pi i L^t_\gamma \int_{M_3} \big(\hat{c}_1^{\rho} \cup \delta \hat{c}_1^{\rho}+b^\rho_2\cup\hat{c}^\rho_1 \big) \bigg).
\end{align}
This is slightly different than the $k=2$ case where there was no analog of the middle term above. Comparing to equation (B.15) of \cite{Choi:2022jqy}, we see indeed that the left-hand side is still a condensation operator and the coefficient in front of the middle term is interpreted as discrete theta angle given by the identity element in $H^3(G_k,U(1))\simeq G_k$ given the coefficient of the middle term.\footnote{More generally, the relevant discrete theta angles of this $G_k$ gauge theory is given by $\mathrm{Hom}\big(\mathrm{Tor}\;\Omega^{Spin}_3,U(1) \big)$ when $M_3$ is a spin manifold. We will leave this more refined consideration of the structure of $M_3$ to future work.}\footnote{Notice that (B.15) of \cite{Choi:2022jqy} is written in terms of $U(1)$ valued forms, where the purpose of their first term is to constrain the gauge field to be discretely valued.} Explicitly we have (restoring the correct normalization),
\begin{align}
     &\mathcal{U}(M_3)\times \mathcal{U}^\dagger(M_3) =\frac{1}{|H_2(M_3,G_k)|}\sum_{\sigma \in H_2(M_3,G_k)} \epsilon(M_3,\sigma) \exp{\bigg(2\pi i L^t_\gamma \int_\sigma b^\rho_2\bigg)}
\end{align}
where $\epsilon(M_3,\sigma)$ is a discrete torsion term and following \cite{Choi:2022jqy}, we see that the right-hand side is equivalent to a level-1 Dijkgraaf-Witten theory with gauge group $G_k$ coupled to a 1-form electric background field $b^\rho_2$. In other words,
\begin{align}
     &\mathcal{U}(M_3)\times \mathcal{U}^\dagger(M_3) =\mathcal{T}^{(1,G_k)}_{DW}(M_3,b^\rho_2)
\end{align}
in the obvious notation. Note that this is again a 3-gauging of a 3-form symmetry along $M_3$.

\end{itemize}



\subsection{More General 6D SCFTs}

We can extend our discussion in a few different directions. One can also consider
more general F-theory backgrounds with constant axio-dilaton \cite{Heckman:2013pva, DelZotto:2014hpa, Heckman:2014qba, Heckman:2015bfa, DelZotto:2017pti}).\footnote{This includes, for example, rank $N$ conformal matter of type $(G,G)$ \cite{DelZotto:2014hpa, Heckman:2014qba}.}
Even though the axio-dilaton is constant, the duality bundle can still be non-trivial.
The base of the model is again a generalized ADE-type singularity, and
has boundary torsional 1-cycles on which we can wrap D3-branes.

More broadly speaking, whenever we have a non-trivial defect group we anticipate that a similar structure persists.
When we have a position dependent axio-dilaton profile at the boundary of the base geometry, it appears simplest to extract the relevant topological terms for the generalized symmetry operators by starting with the topological terms of an M5-brane and dimensionally reducing along the (torsional) 3-cycle obtained by fibering the F-theory torus over the torsional 1-cycle of the base.

It is also natural to treat the effects of the 0-form and 1-form symmetries by explicitly tracking the profile of flavor 7-branes in the system. One way to proceed is to pass to the M-theory limit by compactifying on a further circle. So long as the generalized symmetry operator does not wrap this circle, we can then analyze these effects in purely geometric terms using \cite{Cvetic:2022imb}. Alternatively, we can use the known structure of topological terms on the tensor branch of these 6D theories to extract the same data from a ``bottom up'' perspective \cite{Apruzzi:2020zot, Apruzzi:2021mlh, Hubner:2022kxr, Heckman:2022suy}.

In both situations, however, the appearance of a non-trivial $SL(2,\mathbb{Z})$ bundle in the F-theory background
is a strong indication that the resulting generalized symmetry operators will have a fusion algebra which is not captured
by a group law. Said differently, we expect that generically, these 6D SCFTs will have non-invertible symmetries which act on 3D defects sourcing a background for $G^{(1)}_k$.

We leave a more systematic analysis of these cases for future work.

\section{M-theory Examples} \label{sec:Mtheory}

Although we have focussed on IIB / F-theory backgrounds, the same
considerations clearly hold more broadly. For example, 5D\ SCFTs
engineered via M-theory on Calabi-Yau canonical singularities can also
support various defects \cite{Albertini:2020mdx, Morrison:2020ool,
Tian:2021cif, DelZotto:2022fnw}. To set notation, let $X$ denote a non-compact
Calabi-Yau threefold which generates a 5D\ SCFT. We can get defects by
wrapping M2-branes and M5-branes on non-compact cycles which extend to the
boundary $\partial X$. The corresponding topological operators are obtained by
wrapping magnetic dual branes on the appropriate cycles:%
\begin{align}
\underset{\text{Line Defect}}{\underbrace{\text{M2 on }\widetilde{M}_{1}%
\times\mathbb{R}_{\geq0}\times \widetilde{\gamma}_{1}}}  & \leftrightarrow
\underset{\text{Gen. Symm. Membrane}}{\underbrace{\text{M5 on }{M}%
_{3}\times\gamma_{3}}}\\
\underset{\text{Surface Defect}}{\underbrace{\text{M5 on }\widetilde{M}_{2}%
\times\mathbb{R}_{\geq0}\times \widetilde{\gamma}_{3}}}  & \leftrightarrow
\underset{\text{Gen. Symm. Defect}}{\underbrace{\text{M2 on }{M}%
_{2}\times\gamma_{1}}}\\
\underset{\text{Wall Defect}}{\underbrace{\text{M5 on }\widetilde{M}_{4}\times
\mathbb{R}_{\geq0}\times \widetilde{\gamma}_{1}}}  & \leftrightarrow
\underset{\text{Gen. Symm. Point}}{\underbrace{\text{M2 on pt}\times\gamma_{3}%
}}.
\end{align}
Reduction of the topological terms on the worldvolume of these branes then produces the
corresponding TFT concentrated on our symmetry operator, see \eqref{eq:WZWM5}.

As a final comment on this example, we note that 5D\ SCFTs sometimes also
enjoy flavor symmetries as realized by various discrete symmetries as well as
\textquotedblleft flavor 6-branes\textquotedblright\ (namely
ADE\ singularities). One can in principle consider wrapping such ``6-branes'' on torsional cycles
of the boundary geometry. This can be viewed as introducing a singular profile for the M-theory
metric in the asymptotic geometry. Wrapping such a 6-brane on a torsional 3-cycle would result
in a generalized symmetry operator for a 0-form symmetry (as it is codimension 1 in the 5D spacetime).
Clearly, this case is a bit more subtle to treat, but it is so intriguing that we leave it as an
avenue to pursue in future work.


\section{Further Generalizations}

So far, we have mainly explained how to lift various ``bottom up'' field theory structures to explicit string constructions. This is
already helpful because it provides us with a machine for extracting the corresponding worldvolume TFT on these generalized symmetry operators, as well as the resulting fusion rules.

But the stringy perspective provides us with even more. For one thing, it makes clear the ultimate fate of these ``topological'' operators once we recouple to gravity. Indeed, once we couple to gravity, $X$ no longer has a boundary, and so all of our wrapped branes will again become dynamical. Moreover, we can also see that in many cases, these generalized symmetries automatically trivialize in compact geometries.

Reinterpreting generalized symmetry operators in terms of wrapped branes also suggests a further ``categorical'' generalization of the standard generalized symmetries paradigm. Indeed, it has been appreciated for some time that at least in type II backgrounds on a Calabi-Yau threefold, the spectrum of topological branes is captured, in the case of the topological B-model by the (bounded) derived category of coherent sheaves and in the mirror A-model by the
triangulated Fukaya category.\footnote{See e.g., \cite{Kontsevich:1994dn, Douglas:2000gi, Aspinwall:2001pu} and \cite{Aspinwall:2004jr} for a review.} The important point here is that for these more general objects, simply working in terms of ``branes wrapped on cycles'' is often inadequate. This in turn suggests that instead of assigning a generalized symmetry operator to a sub-manifold of the $d$-dimensional spacetime, it is more appropriate to work in terms of a complex of objects (in the appropriate derived category). Note also that because these derived categories are monoidal, there is also a notion of fusion in this setting.


\section*{Acknowledgements}

We thank M. Del Zotto, S. Nadir Meynet and R. Moscrop for helpful discussions,
and I. Garcia Etxebarria for a helpful question on an earlier version of this paper.
The work of JJH and ET is supported by DOE (HEP) Award DE-SC0013528. The work
of MH and HYZ is supported by the Simons Foundation Collaboration grant
\#724069 on ``Special Holonomy in Geometry, Analysis and Physics''.



\bibliographystyle{utphys}
\bibliography{TopOp}

\end{document}